\documentclass{article}

\usepackage[creativeai, preprint]{neurips_2025}

\usepackage[utf8]{inputenc} 
\usepackage[T1]{fontenc}    
\usepackage{hyperref}       
\usepackage{url}            
\usepackage{booktabs}       
\usepackage{amsfonts}       
\usepackage{nicefrac}       
\usepackage{microtype}      
\usepackage{xcolor}         
\usepackage[pdftex]{graphicx}
\usepackage{wrapfig}

\usepackage[dvipsnames]{xcolor}

\title{BLK-Assist: A Methodological Framework for Artist-Led Co-Creation with Generative AI Models}

\author{%
  Daniel Grimes
  \\
  Visual-Synthesizer.com \\
  Grass Valley, CA 95945 \\
  \texttt{danieljoygrimes@gmail.com} \\
  \And
  Rachel M.~Harrison \\
  Ophiuchus LLC\\
  Dover, DE 19904\\
  \texttt{rae@ophiuchus.ai} \\
}

\begin{document}

\maketitle

\begin{abstract}
  This paper presents BLK-Assist, a modular framework for artist-specific fine-tuning of diffusion models using parameter-efficient methods. The system is implemented as a case study with a single professional artist’s proprietary corpus and consists of three components: BLK-Conceptor (LoRA-adapted conceptual sketch generation), BLK-Stencil (LayerDiffuse-based transparency-preserving asset generation), and BLK-Upscale (hybrid Real-ESRGAN and texture-conditioned diffusion for high-resolution outputs). We document dataset composition, preprocessing, training configurations, and inference workflows to enable reproducibility with publicly available models to illustrate a privacy-preserving, consent-based approach to human–AI co-creation that maintains stylistic fidelity to the source corpus and can be adapted for other artists under similar constraints.
\end{abstract}

\begin{wrapfigure}{r}{0.45\linewidth}
    \vspace*{-0.15in}
    \centering
    \includegraphics[width=.95\linewidth]{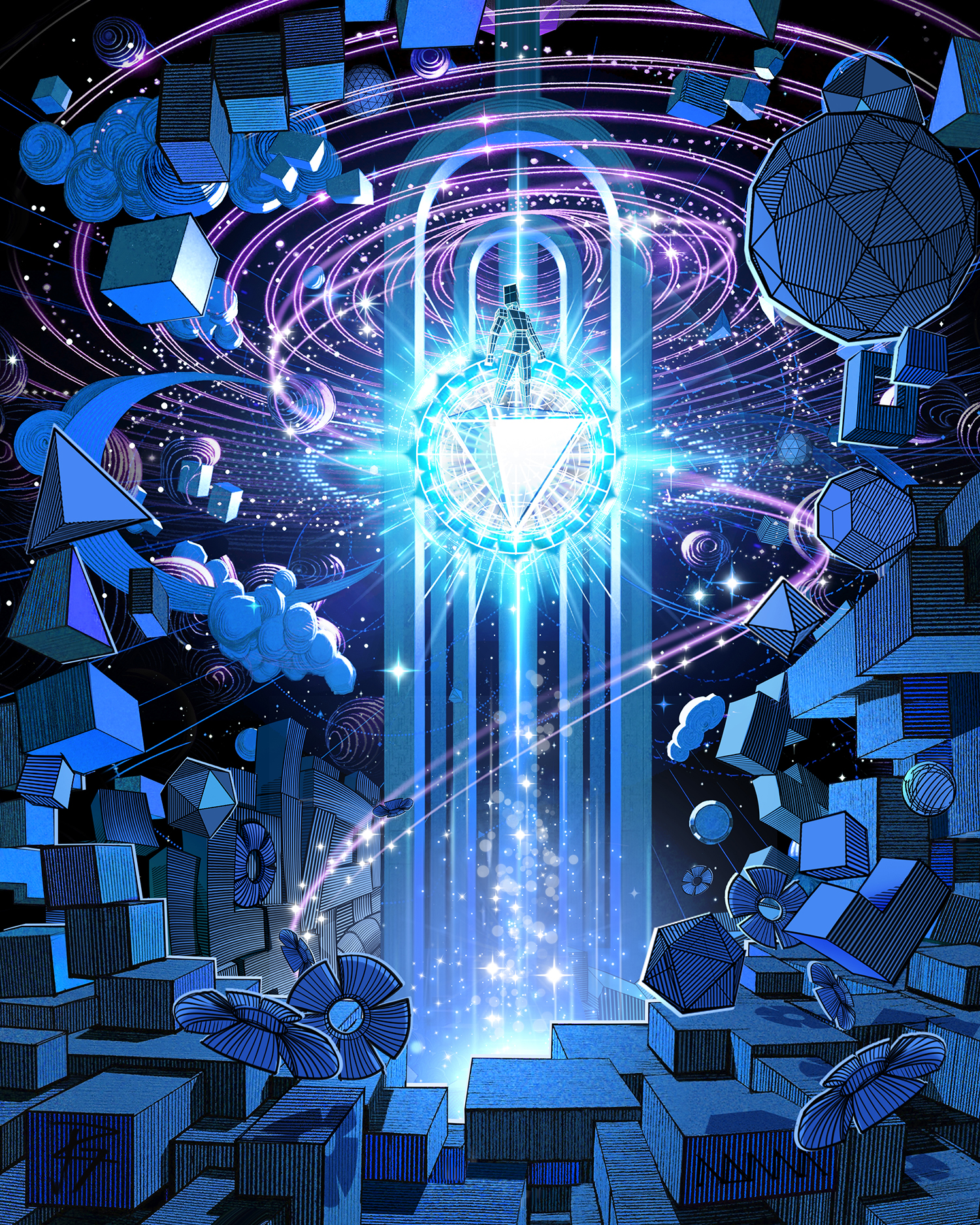}
    \caption{\emph{The Throat Shakra} (2025),  produced with the BLK-Assist pipeline.}
    \vspace*{0.1in}
\end{wrapfigure}

\section{Introduction}

The rapid advancement of large-scale text-to-image (TTI) models has ignited significant debate across creative industries.
On one hand, critics argue that these systems are built on non-consensual data scraping that raises profound concerns around copyright, attribution, and the displacement of artistic labor~\cite{jiang2023ai, steinbruck2023creative, oppenlaender2023text}.
On the other, advocates  counter that they act as a democratizing force to lower technical barriers and enable practitioners of all skill levels to produce complex visual works via prompt engineering~\cite{oppenlaender2022creativity, ko2023large}.
However, this paper proposes a third, artist-centric paradigm: the AI Studio Assistant.
Rather than acting as an autonomous creator or generic tool, the assistant becomes a modern-day protégé personalized by and for an individual artist through fine-tuning on their own proprietary corpus to augment their workflows while preserving authorship and agency in a private and controlled environment.

This concept builds on a longstanding lineage of co-creation in the arts.
Renaissance masters like Raphael and Rubens relied on workshops where assistants executed parts of a composition under the master’s direction, while Andy Warhol’s \textit{The Factory} reframed the artist’s role as a creative director orchestrating an assembly-line process \cite{lisocietal}.
In each case, authorship was defined by conceptual vision rather than fully manual execution--a definition that
also aligns with U.S. Supreme Court interpretations of authorship in copyright law, which have historically defined an author as "he to whom anything owes its origin"~\cite{burrowgiles1884, usco2023aiRegistration}.

The foundation of this paradigm rests on principles of personalization, agency, and ethics, and is supported by established theories in Human-Computer Interaction (HCI)~\cite{preece1994human, mackenzie2024human, dix2009human}.
Research in human–computer co-creativity shows that embedding an artist’s style and working methods directly into a generative model increases both agency and ownership \cite{abuzuraiq2024towards}.
Such results can often be achieved through adopting a "small data" mindset, which advocates for fine-tuning models on carefully curated, artist-owned datasets rather than relying on massive, indiscriminately scraped web data~\cite{abuzuraiq2024towards, vigliensoni2022small}.

As part of this effort, we present \textbf{BLK-Assist}: a modular generative AI Studio Assistant pipeline built on publicly available diffusion models fine-tuned through parameter-efficient fine-tuning (PEFT)~\cite{han2024parameter} on a singular artist's own curated dataset.
BLK-Assist consists of three interoperable models:

\begin{itemize}
    \item \textbf{BLK-Conceptor} is a LoRA-adapted diffusion model for rapid conceptual sketching, composition, and lighting exploration. It generates varied ideas in minutes while preserving the artist’s signature style for continuity across storyboards, mood studies, and concept art.

    \item \textbf{BLK-Stencil} is a transparency-preserving asset generator for modular, collage-style composition. It produces layered PNG assets that integrate with BLK-Conceptor sketches for a seamless concept-to-composition workflow.

    \item \textbf{BLK-Upscale} is a hybrid super-resolution pipeline trained on ultra-high-res scans for gallery-quality output. Real-ESRGAN handles small to medium prints, while large-scale works (up to 3B px) use a custom Diffusion+Texture LoRA with recursive tiling to preserve fine brushwork and surface detail at mural and exhibition scales.
\end{itemize}

By detailing the data, training configurations, and overall workflows for each component, we offer a replicable method for other artists and researchers to engage with a new approach to human-AI co-creation that preserves artistic agency, avoids the ethical pitfalls of large-scale data scraping, and respects the intimacy of individual human expression by allowing both creativity and authorship to rest firmly within the artist's design of the interactive system itself.

\section{Methodological Framework}

BLK-Assist is a modular framework consisting of three separate models designed to augment disparate steps in the artistic process.
Though the models themselves are not publicly released, we present a detailed account of the training data, architecture selection, preproccessing, and hyperparameter tuning to facilitate reproducibility by independent researchers using open tools.

\begin{figure}
    \centering
    \includegraphics[width=\linewidth]{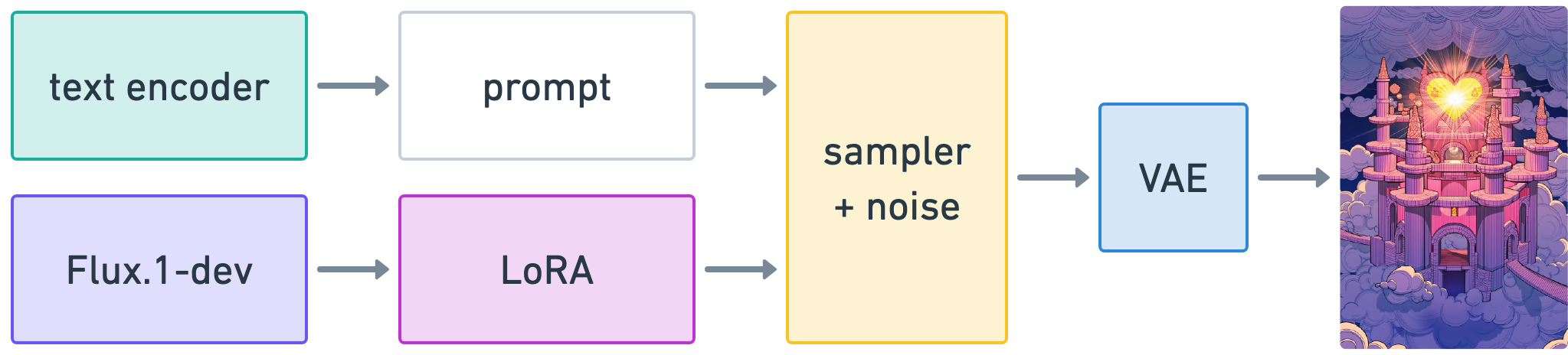}
    \caption{Overview of the BLK-Conceptor inference pipeline.}
    \label{fig:BLK-Conceptor}
\end{figure}

\subsection{BLK-Conceptor}

BLK-Conceptor (Figure~\ref{fig:BLK-Conceptor}) was designed to generate conceptual sketches for quick ideation. 
It was fine-tuned on Black Forest Labs \texttt{Flux.1-dev}~\cite{flux2024, labs2025flux1kontextflowmatching} using LoRA~\cite{hu2022lora}, which was chosen for its parameter-efficient training ability and ease of use in any inference pipeline.

\begin{itemize}
    \item \textbf{Input/Output:} Text-to-image generation at any aspect ratio, with best performance \textasciitilde1 Mpx (e.g., $1024\times1024$) during sampling.
    Prompts were curated for content generalization, lighting, mood, and accurate descriptions of the artwork. During training, image captions were provided via vLLM~\cite{kwon2023efficient} with few-shot examples in the system prompt, then iteratively refined (see Figure~\ref{fig:blk_conceptor_outputs}. Data pre-processing focused on full compositions, with many iterations converging on a $72/25$ full-composition/detail-shot ratio as ideal.
    \item \textbf{Training dataset:} The curated training dataset consisted of hand-drawn pen-and-ink and digital illustrations emphasizing full compositions, targeting an approximate $72/25$ full-composition/detail ratio due to strong composition abilities and good detail for the target resolution of $1024\times1024$ (\textasciitilde1 million pixels). Captions were stored as sidecar \texttt{.txt} files. See Figure~\ref{fig:blk_conceptor_inputs} for examples of training data.
    \item \textbf{Architecture:} This project used \texttt{Flux.1-dev}~\cite{flux2024, labs2025flux1kontextflowmatching} with a LoRA network~\cite{hu2022lora} (rank 32, linear\_alpha 32) attached to U-Net~\cite{ho2020denoising} with a frozen text encoder. Adapters were trained with fixed base model weights. Training was done with 
    8-bit linear weight quantization with Optimum Quanto~\cite{optimumquanto2025} enabled for efficiency.
    \item \textbf{Training approach:} Supervised fine-tuning (SFT)~\cite{ouyang2022training} with a flow-matching noise scheduler (rectified-flow style objective) inference. Human-in-the-loop evaluation guided prompt/caption revisions and checkpoint selection.
    \item \textbf{Hyperparameters:} \textbf{Steps:} 4,000-8,000 (saving every 250; EMA decay 0.99; keep 16 checkpoints); \textbf{Batch size / Accumulation:} 1 / 1; \textbf{Optimizer:} Prodigy (lr 1.0; weight\_decay 0.001; decoupled; bias correction; warmup safeguard; d0=1e-5, d\_coef=2.0); \textbf{Precision:} bf16 (training), float16 (saves); \textbf{Trainable modules}: \texttt{UNet=true}, \texttt{Text-encoder=false}, Grad checkpointing: true
    \item \textbf{Additional preprocessing:} Multi-resolution training at 512, 768, and 1024 pixels; caption\_dropout 0.05; token shuffling enabled; latents cached to disk for throughput.
\end{itemize}

\begin{figure}
    \centering
    \includegraphics[width=\linewidth]{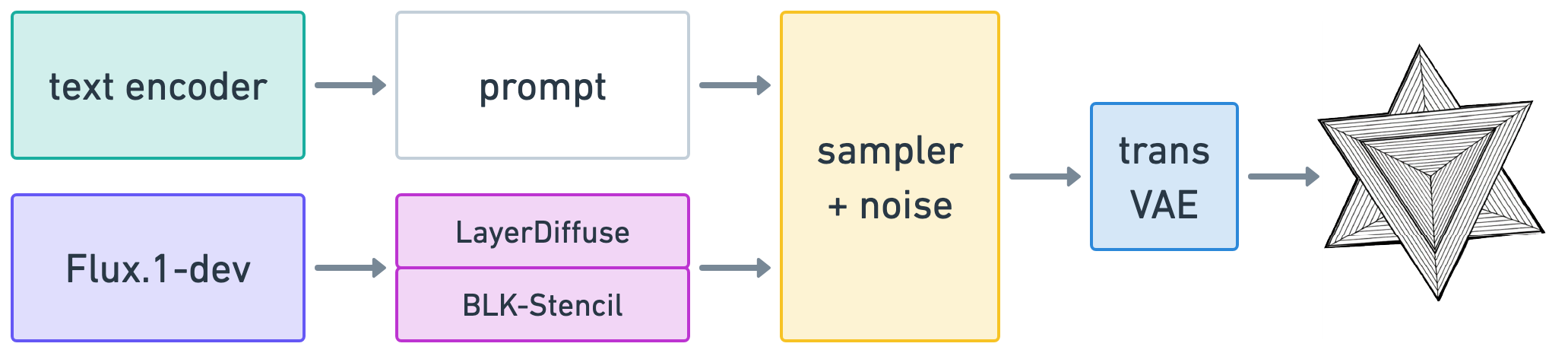}
    \caption{Overview of the BLK-Stencil inference pipeline.}
    \label{fig:BLK-Stencil}
\end{figure}

\subsection{BLK-Stencil}

BLK-Stencil (see Figure~\ref{fig:BLK-Stencil}) is an RGBA-layered asset generator combining \texttt{Flux.1-dev}, LayerDiffuse transparency conditioning~\cite{zhang2024transparent, redaigcfluxlayerdiffuse}, and LoRA adapters~\cite{hu2022lora}. It was designed to produce transparent PNG elements for modular composition, enabling foreground, midground, and background overlays to be assembled with precise stylistic and spatial control. This work builds on the transparency-preserving approach described by Zhang et al.~\cite{zhang2024transparent}, with early prototyping using the SDXL implementation~\cite{lllyasvielsdforgelayerdiffuse}
and final development based on the Flux adaptation
~\cite{redaigcfluxlayerdiffuse}.

\begin{itemize}
    \item \textbf{Asset taxonomy:} The core dataset comprised Platonic solids (tetrahedron, cube, octahedron, dodecahedron, icosahedron) in both single-object form and multi-object arrangements, which were then further diversified by line weight (thin, medium, thick). Additional style assets included stylized clouds, and other geometric shapes (see Figure~\ref{fig:blk_stencil_inputs}. Each asset was tagged with a Z-role (FG, MG, BG-overlay) for compositional placement.
    \item \textbf{Dataset \& captions:} The dataset followed a hierarchical folder structure by category, with captions stored as sidecar \texttt{.txt} files alongside each PNG. Caption templates specified style, subject, layering, and rendering details. All PNGs preserved alpha channels.
    Data preprocessing used examples of platonic solids and other hand drawn digital assets with transparent alpha-layer conditioning; the platonic solid data was synthetically generated in 3D, then hand-drawn over to achieve a perfect stylistic match. See Figure~\ref{fig:blk_stencil_inputs} for examples.
    \item \textbf{Augmentation:} Alpha-safe augmentations included flips, $\pm$10--15\% rotation/scale, light jitter, and mild grain. Alpha-smearing operations and premultiplied blurs were avoided. For multi-object composites, alpha-aware compositing ensured clean layer boundaries.
    \item \textbf{Training setup:} The base model used \texttt{Flux.1-dev}~\cite{flux2024, labs2025flux1kontextflowmatching} with a frozen LayerDiffuse LoRA and transparent VAE for alpha-channel preservation~\cite{lllyasvielsdforgelayerdiffuse, redaigcfluxlayerdiffuse}.
    A LoRA~\cite{hu2022lora} adapter was trained for style transfer. 
    The initial rank was $r=16$, but increased to $r=32$ for intricate structures. Optimization used AdamW~\cite{loshchilov2017decoupled} 8-bit with learning rate $2\times10^{-6} \rightarrow 5\times10^{-6}$ under a polynomial decay schedule, decreasing progressively as training advanced to stabilize convergence. Training followed a curriculum: singles (1--2k steps), then +pairs (40\% of dataset; 2--6k steps), then +triples (20\%) with extras mixed in ($>$6k steps).
    \item \textbf{Prompting recipes:} Prompts included both asset descriptors and Z-role tags (see Figure~\ref{fig:blk_stencil_outputs}).

\end{itemize}

\begin{figure}
    \centering
    \includegraphics[width=\linewidth]{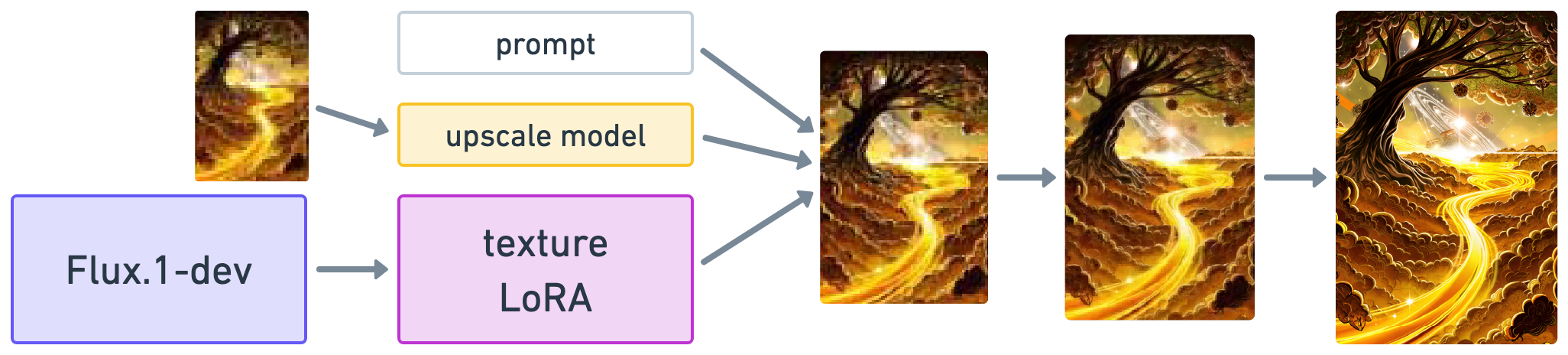}
    \caption{Overview of the BLK-Upscale Diffusion+Texture LoRA inference pipeline.}
    \label{fig:BLK-Upscale}
\end{figure}

\subsection{BLK-Upscale}

\begin{wrapfigure}{r}{0.45\linewidth}
    \vspace*{-0.15in}
    \centering
    \includegraphics[width=.95\linewidth]{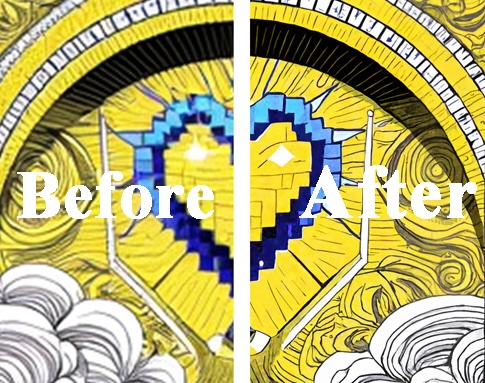}
    \caption{A before and after example of low-res artwork upscaled with BLK-Upscale.}
    \vspace*{-0.2in}
    \label{fig:blk_upscale_example}
\end{wrapfigure}

BLK-Upscale (see Figure~\ref{fig:BLK-Upscale}) is a hybrid super-resolution system that allows artist to select between two complementary upscaling paths based on target size and desired texture fidelity. For modest enlargements, a fine-tuned Real-ESRGAN~\cite{wang2021real} preserves linework and tone with minimal artifacts. For mural and gigapixel outputs, a diffusion path employs a texture-focused LoRA with tiled refinement to synthesize convincing hand-rendered microstructure at extremely high resolutions.

Real-ESRGAN is intended for faithful 4x--8x enlargement for quality fast upscales approaching 8k resolution. Diffusion+Texture LoRA is preferred when target resolution exceeds 8k, or where controlled texture synthesis needs to be convincing at large scale.

\paragraph{Real-ESRGAN:}
Fine-tuned \texttt{Real-ESRGAN\_x4plus}~\cite{wang2021real} on finished paintings and scans. Best for 4x--8x global upscales where faithful line preservation and tone stability dominate.

\begin{itemize}
    \item \textbf{Input/Output:} Flattened RGB artworks (digital paintings and HR scans). Outputs are PNGs at target resolution with no alpha layers. Input/Output example can be seen in Figure~\ref{fig:blk_upscale_gan}.
    \item \textbf{Training Datasets:} Three $10,000\times10,000$ px 8-bit PNG artworks. During training, $256\times256$ HR patches were dynamically sampled with a detail-weighted probability (Laplacian magnitude) and optional artist-defined masks. Corresponding $64\times64$ LR patches (4x scale) were generated on-the-fly via the Real-ESRGAN degradation pipeline~\cite{wang2021real}—kernel blur, noise, downsampling, and JPEG compression. Continuous sampling provides infinite augmentation.
    \item \textbf{Architecture:} Generator: RRDBNet (3$\rightarrow$3, 64 feat, 23 RRDB, grow=32, scale=4); Discriminator: UNetDiscriminatorSN. Initialized from \texttt{RealESRGAN\_x4plus}~\cite{wang2021real}.
    \item \textbf{Training Approach:} On-the-fly HR crops (gt\_size 192--256). Loss mix: L1 (1.0) + VGG19 perceptual pre-activation (0.1--0.5) + relativistic average GAN (0.005--0.01). Augment: flips, 90$^{\circ}$ rotations, light brightness/contrast jitter. Adam (lr $1\times10^{-4}$, betas 0.9/0.999), multi-step decay at 40k/80k. 50k--100k iterations total with early stopping by visual QC.
    \item \textbf{Inference and Resolution Stepping:} Global 2x--4x upscale, with very large canvases using tiled inference (tile 512, pad 16--32) to minimize seams.
\end{itemize}

\paragraph{Diffusion+Texture LoRA:}
Flux-based tiled refinement using a style-specific texture LoRA. Run via ComfyUI Ultimate SD Upscale~\cite{comfyui_ultimatesdupscale} with controlled denoise and overlap; applied in 2x--4x passes to inject hand-drawn microtexture consistently across tiles.

\begin{itemize}
    \item \textbf{Input/Output:} Flattened RGB artworks (digital paintings and HR scans). Outputs are PNGs at target resolution with no alpha layers. See Figures~\ref{fig:blk_upscale_example} and \ref{fig:blk_upscale_diffusion} for examples.
    \item \textbf{Training Datasets:} Datasets consisted of 1024--1536 px crops of texture primitives from finished pieces and scans (ink hatch, cross-hatch, paper grain, brush streaks). 300--2000 crops with short, texture-centric captions in sidecar \texttt{.txt} files.
    \item \textbf{Architecture:} \texttt{Flux.1-dev}~\cite{flux2024, labs2025flux1kontextflowmatching} backbone with LoRA adapters on attention (\texttt{q/k/v/out}; optional FF proj). Texture LoRA rank $r=8$--32 (task-dependent). Ultimate SD Upscale node for tiled diffusion.
    \item \textbf{Training Approach:} LoRA adapters were trained on the \texttt{Flux.1-dev}~\cite{flux2024, labs2025flux1kontextflowmatching} model. \textbf{Adapter:} \texttt{LoRA} with rank $r=16$ and \texttt{alpha}=16; target modules: \texttt{to\_q}, \texttt{to\_k}, \texttt{to\_v}, \texttt{to\_out.0}; \texttt{bias=none}, \texttt{dropout=0.0}; \textbf{Precision \& accumulation:} \texttt{bf16} training, gradient accumulation $=$ 8; \textbf{Optimizer:} \texttt{adamw\_8bit} with learning rate $5\times10^{-6}$, \texttt{betas}=(0.9, 0.999), \texttt{weight\_decay}=0.01; \textbf{LR schedule:} \texttt{cosine} decay with warmup of 500 steps (about 1--3\% of total), total scheduled steps $=$ 10{,}000; \textbf{Training length:} \texttt{max\_train\_steps}=10{,}000.
    \item \textbf{Inference and Resolution Stepping:} Multi-pass 2x--4x steps: Pass A injects texture (denoise 0.25--0.37, LoRA 1.0), Pass B refines (0.17--0.23, LoRA 1.0), and optional Pass C stabilizes edges (0.13--0.17, LoRA 1.0). Tile size 512--1536 pixels with 64--128 pixel overlap; seam blend on. Final low-denoise full-frame pass at target res for global coherence. Diffusion path can be repeated as many times as desired but becomes increaingly computationally expensive and hyperparameters/prompts must be tuned to match tile sizes and intent. Often, prompts are left empty and system relies on flow matching in context (IC) knowledge of texture LoRA for texture logic. 
\end{itemize}

\subsection{Computational Environment}

All models were trained on a custom-built AI assistant workstation featuring a 24-core AMD Threadripper CPU, 256 GB of RAM, dual NVIDIA RTX A6000 GPUs, and dual NVIDIA RTX 3090 GPUs, with a RAID 1 array and high-speed SSD storage and SSD swap drives for fast dataset pre-processing. The system runs Arch Linux, with Python environments managed via pyenv~\cite{pyenv2025} and Conda~\cite{conda_contributors_conda_A_system-level}. 
Training and experimentation were conducted using ComfyUI~\cite{comfyUI2025}, Automatic1111~\cite{AUTOMATIC1111_Stable_Diffusion_Web_2022}, AI Toolkit~\cite{aitoolkit2025}, SimpleTuner~\cite{simpletuner2025}, and Hugging Face diffusers~\cite{von_Platen_Diffusers_State-of-the-art_diffusion}, each executed with the Python versions and dependencies specified by the respective tools.

All training data was sourced from the personal archive of Adam Dryden\footnote{\url{https://adamdryden.com/}}, curated by Adam Dryden and Daniel Grimes, and comprises a total of 75 full images and 25 details shots for the BLK-Conceptor model, 101 images for the BLK-Stencil model, 125 images for the BLK-Upscale Diffusion+Texture LoRA model, and a streaming patch-based dataset for the BLK-Upscale SR-GAN model, where training samples are generated on the fly by the dataloader.  

Following baseline tests, we applied data augmentation and found horizontal flipping to be an effective method for doubling the dataset size and improving model output quality. A sophisticated prompt design language employing vLLM system prompting with few-shot examples was used for captioning and iteratively refined.
A human-in-the-loop process was integral at every stage to ensure consistent alignment with the intended artistic standards, vision, and interpretation.

Given the highly personal nature of many artistic projects, privacy is often a core operational requirement. This influenced both the hardware (favoring local compute over cloud resources) and design of BLK-Assist’s workflow to ensure all sensitive data remained within a secure environment.

\section{Discussion}
\label{discussion}

The AI Studio Assistant system illustrated in the BLK-Assist framework has significant implications for artistic practice, copyright, and the broader discourse on co-collaborative HCI with generative AI.
In structuring the generative process around an individual artist’s own corpus, the framework demonstrates a concrete, technically reproducible pathway for building personalized systems that preserve authorship, integrate ethical safeguards, and operate within a clearly defined scope of creative agency.
This paradigm also fundamentally reframes the discourse from “deskilling” to “reskilling,” as the artist’s craft expands beyond manual execution to encompass new competencies in data curation, model tuning, and creative direction. Agency is thus expressed not just in the final composition, but in the architectural design of the creative system itself, as the focus of labor shifts from executor to director and the personalized generative model becomes a primary artifact of the artist’s skill.

Rather than positioning AI as an autonomous creator, BLK-Assist treats it as an embedded component of the artist’s workflow that's subject to their direction, constraints, and aesthetic standards, thereby reframing generative AI as an instrument of practice rather than a source of authorship.

\subsection{Ethical Considerations}
The central challenge to the copyrightability of AI-generated art is the requirement of human authorship. The U.S. Copyright Office has clarified that the use of AI as an assistive tool is permissible, provided the human author exercises sufficient creative control over the final work~\cite{usco2025part2}, emphasizing prior determinations that "copyright can protect only material that is the product of human creativity"~\cite{usco2023aiRegistration, uscopyright-faq-protect}. The BLK-Assist workflow provides a strong case for such human authorship by demonstrating creative control at multiple, decisive stages: the curation of the training data, the design of the fine-tuning process, the critical selection of AI-generated components, and the final manual execution of the artwork.

Furthermore, by training exclusively on the artist's own consented work, this model entirely sidesteps the contentious legal and ethical debates surrounding fair use and the non-consensual scraping of copyrighted data that underpin lawsuits against large-scale model developers. While the potential exists for these methods to be exploited to replicate another artist’s style without consent, the scarcity of sufficiently high-quality, non-consenting datasets presents a practical barrier to such misuse.

This approach has the additional benefit of increasing access and accessibility to the act of creation, particularly for artists facing physical disabilities or the limitations of aging. By automating the most physically demanding and repetitive aspects of art production, the framework allows an artist's role to shift from one of manual executor to creative director to preserve their accumulated skill and aesthetic judgment while providing a sustainable path for continued creative practice when physical barriers might otherwise make it impossible.

\subsection{Limitations \& Future Work}
This paper presents a methodological framework based on a single-artist case study. As such, its primary limitation is the lack of a broad, quantitative evaluation, and future work should involve applying this framework to a diverse range of artists and artistic styles to validate its generalizability.
Further research could also develop qualitative and quantitative metrics to assess the workflow by measuring creative satisfaction, perceived agency, and the novelty of the co-created outputs.

On the technical side, several areas could further improve the underlying model pipelines.
Initial testing suggests that LyCORIS LoKR~\cite{yeh2024navigating}
performs well for multi-concept datasets in BLK-Conceptor, with behavior resembling full-rank fine-tuning but requiring lower learning rates and longer schedules.
LoKR also demonstrated improved prompt accuracy and reduced token bleeding, likely due to its enhanced matrix representation capacity.
A next step is to evaluate LoKR comprehensively across all three BLK-Conceptor, BLK-Stencil, and BLK-Upscaler pipelines.

Higher-precision, non-quantized training has shown promising fidelity and stability gains over quantized approaches, and Flux Kontext~\cite{labs2025flux1kontextflowmatching} shows potential for integration into upscale workflows and warrants targeted fine-tuning trials. Future work will also explore fine-tuning on newer non-distilled architectures like Qwen-Image~\cite{wu2025qwenimagetechnicalreport}, Wan 2.2~\cite{wan2025}, and HiDream~\cite{hidreami1technicalreport}, and incorporate AI-assisted high-quality outputs into training cycles to expand datasets and diversify styles.

\bibliographystyle{plain}
\bibliography{references}






\newpage
\appendix

\section{Technical Appendices and Supplementary Material}

\begin{figure}[!htb]
    \centering
    \includegraphics[width=\linewidth]{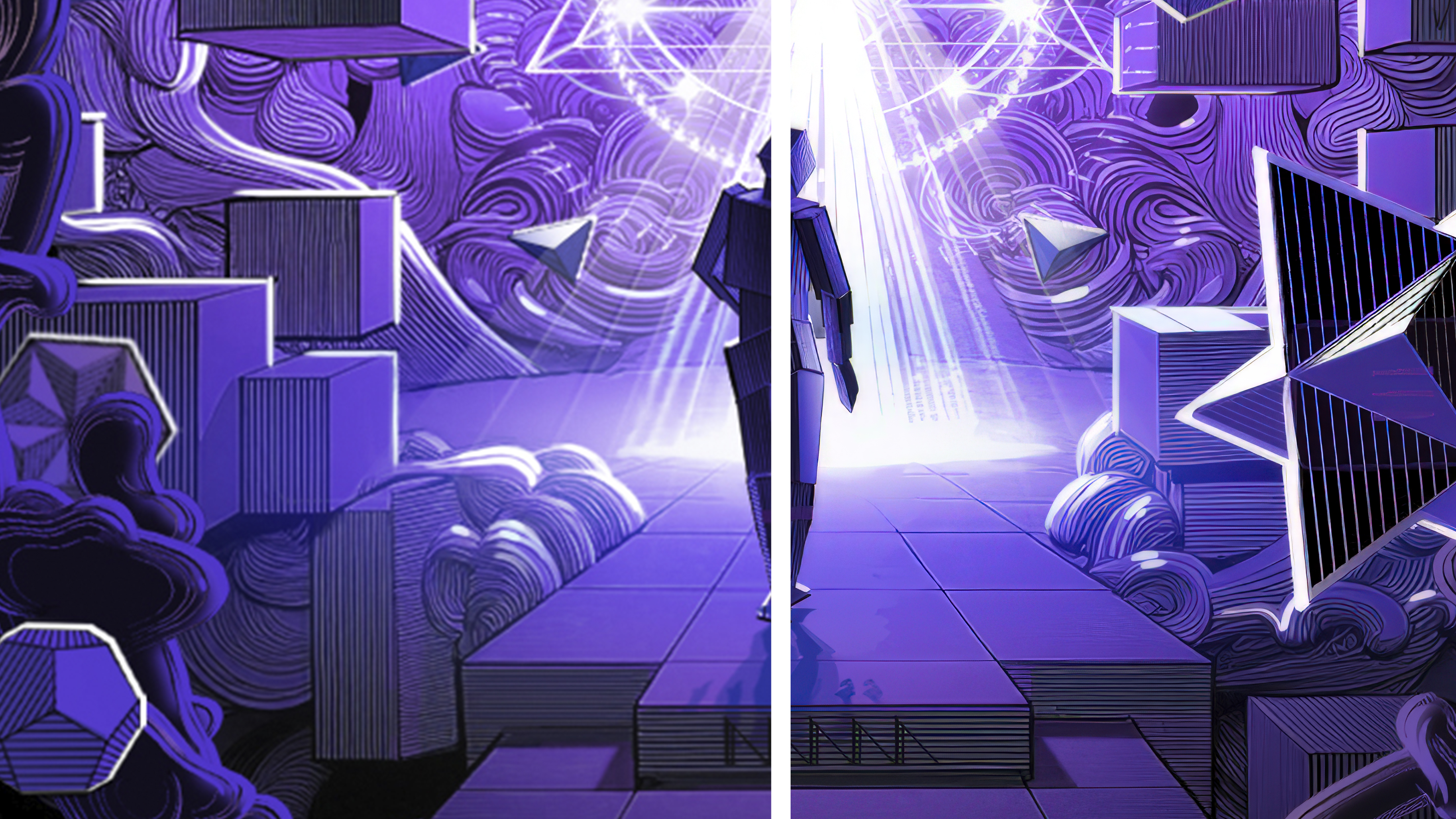}
    \\
    \caption{A before and after example of a low-res image upscaled with BLK-Upscale Real-ESRGAN}.
    \label{fig:blk_upscale_gan}
\end{figure}

\begin{figure}[!htb]
    \centering
    \includegraphics[width=\linewidth]{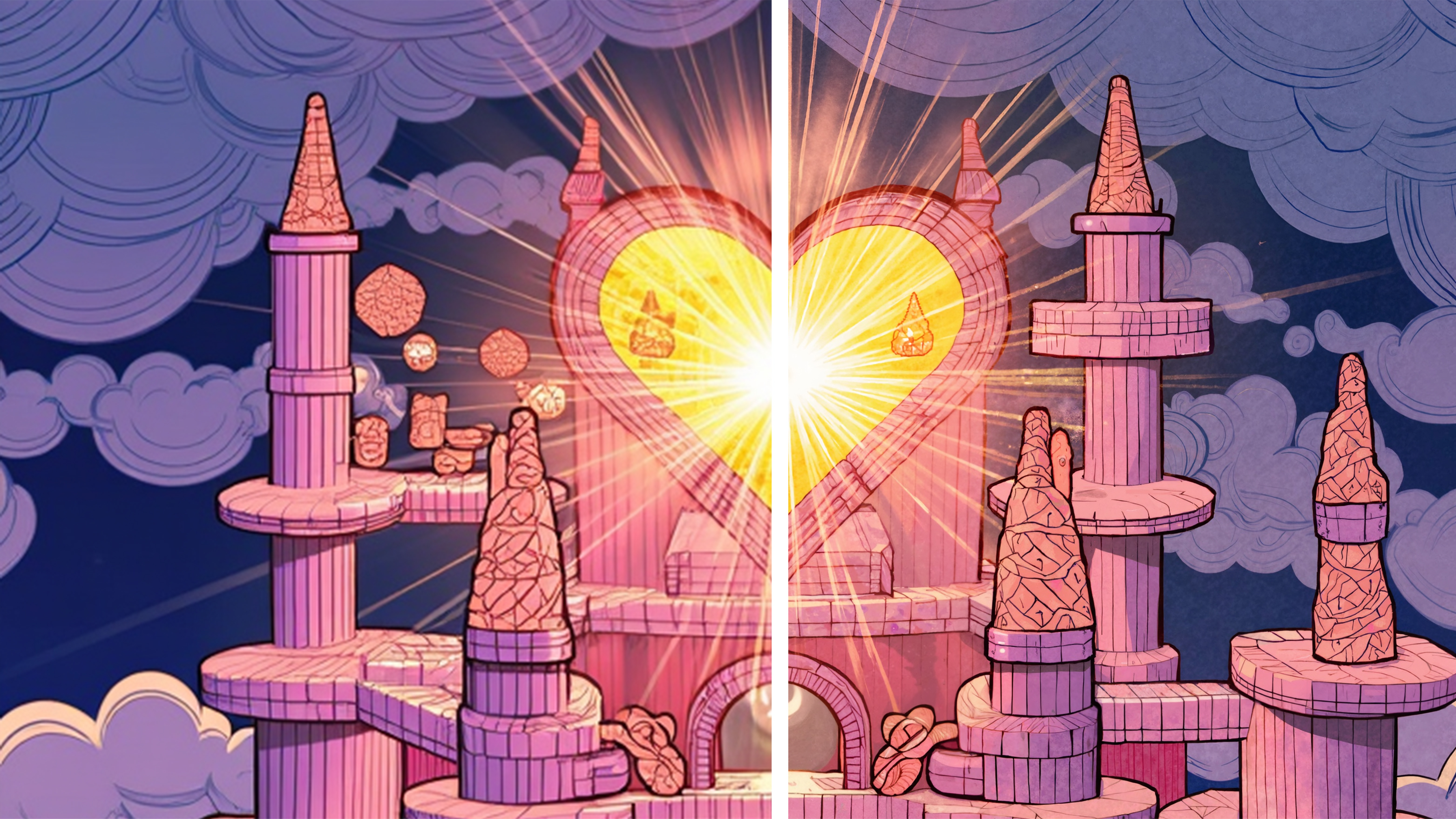}
    \\
    \caption{A before and after example of a low-res image upscaled with BLK-Upscale Diffusion+Texture LoRA}.
    \label{fig:blk_upscale_diffusion}
\end{figure}

\begin{figure*}
    \centering
    \includegraphics[width=\linewidth]{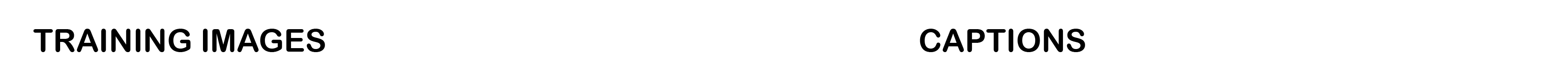}
    \\
    \includegraphics[width=\linewidth]{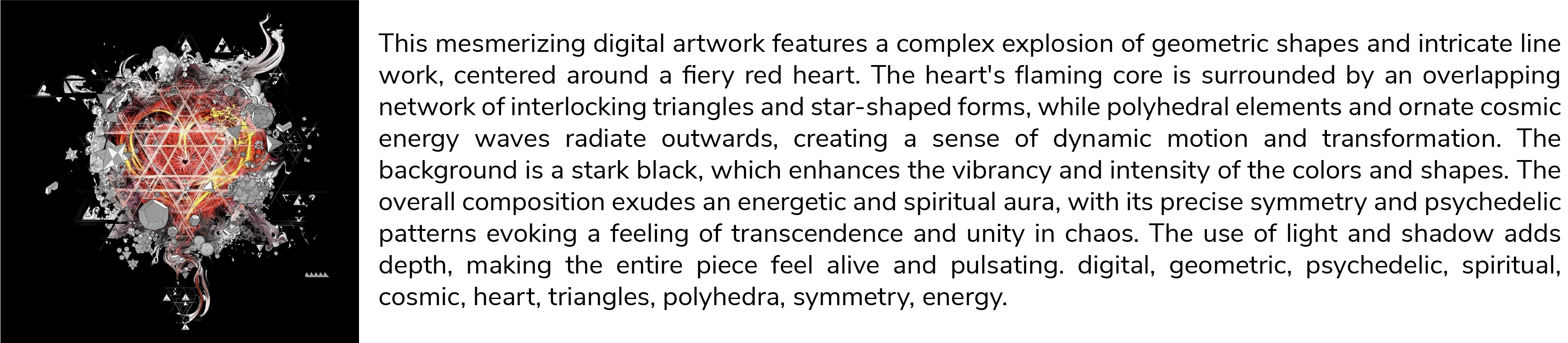}
    \\
    \includegraphics[width=\linewidth]{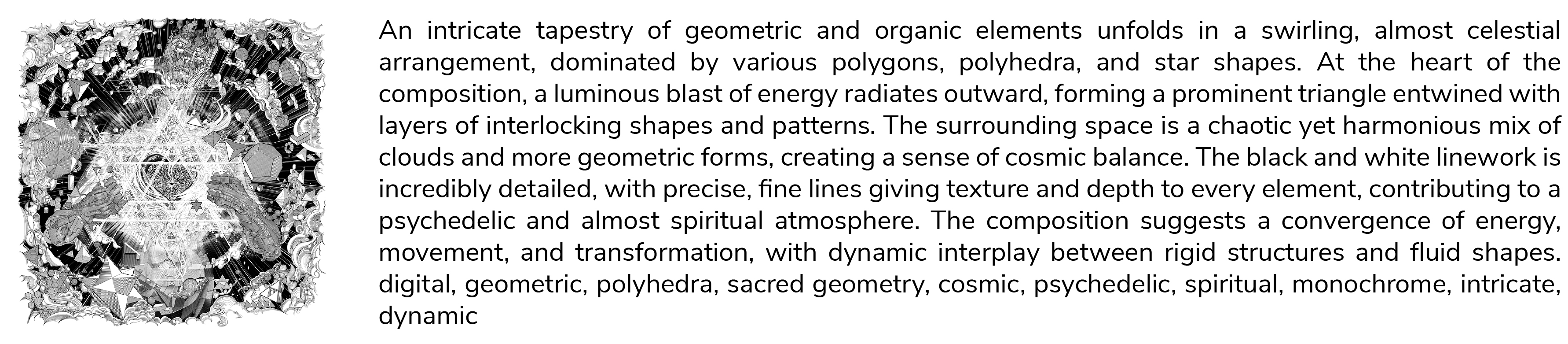}
    \\
    \includegraphics[width=\linewidth]{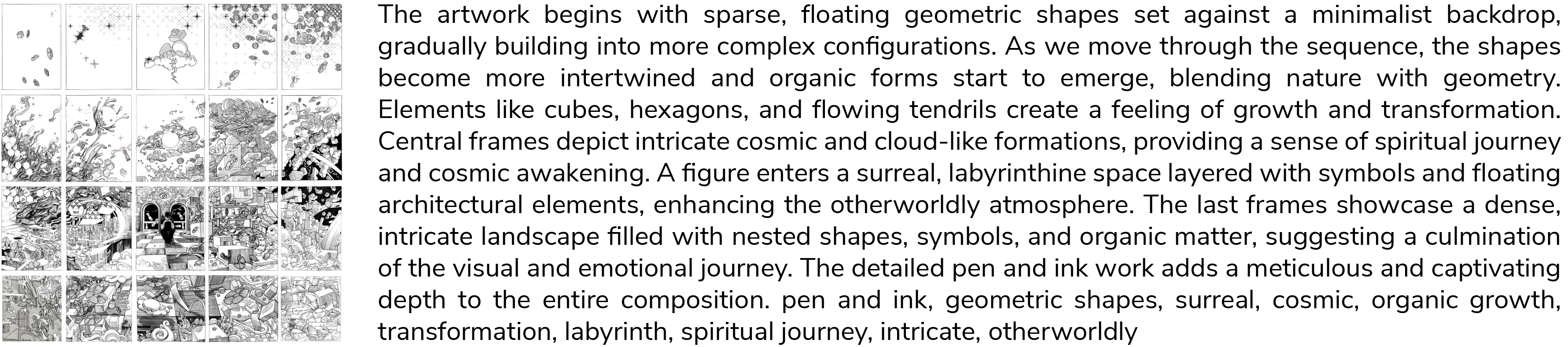}
    \\
    \includegraphics[width=\linewidth]{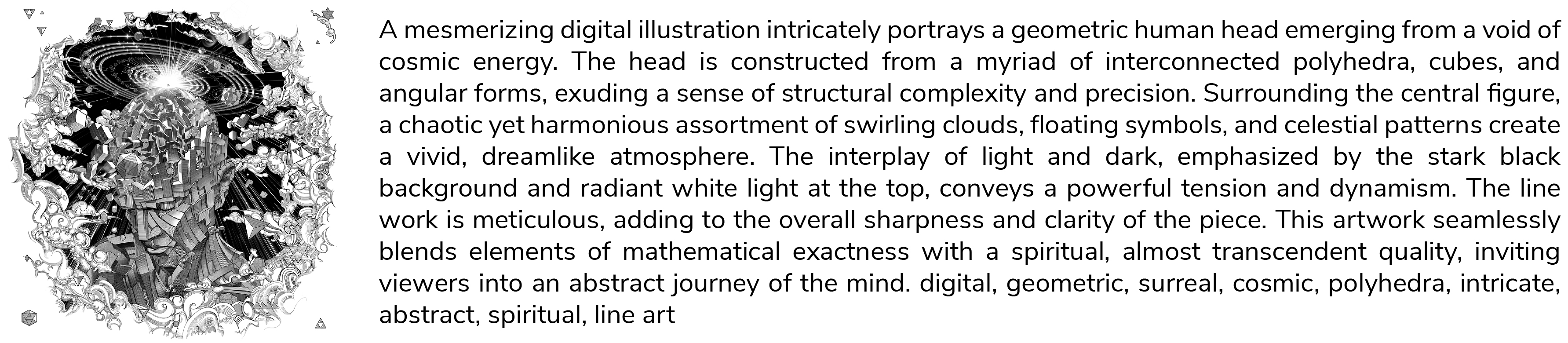}
    \\
    \includegraphics[width=\linewidth]{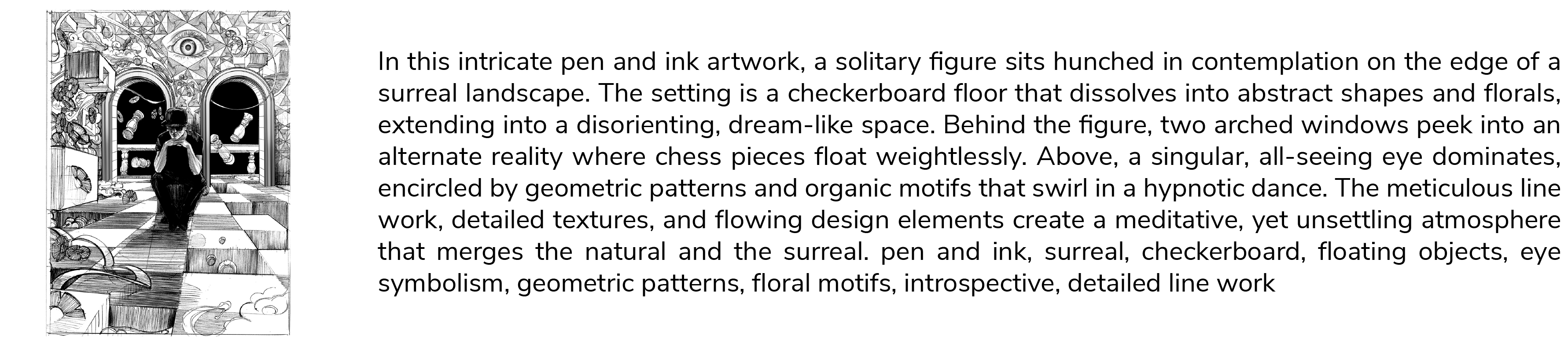}
    \caption{Input training data (images and captions) for BLK-Conceptor}.
    \label{fig:blk_conceptor_inputs}
\end{figure*}

\begin{figure*}
    \centering
    \includegraphics[width=\linewidth]{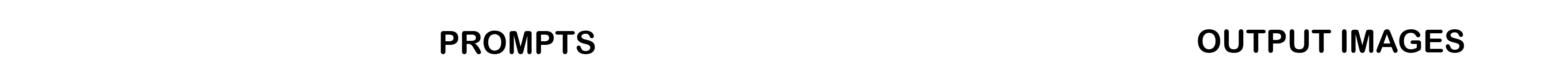}
    \\
    \includegraphics[width=\linewidth]{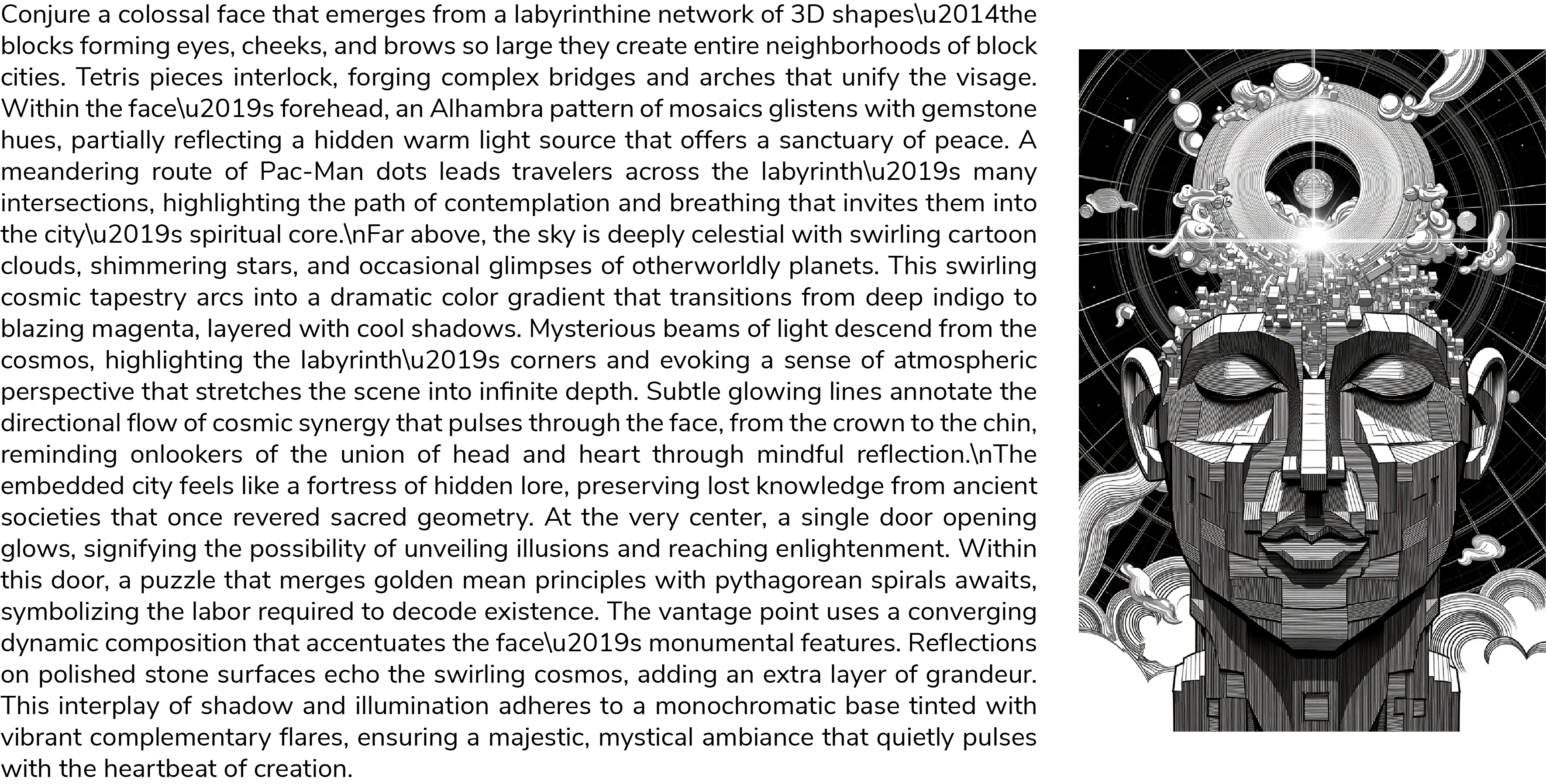}
    \includegraphics[width=\linewidth]{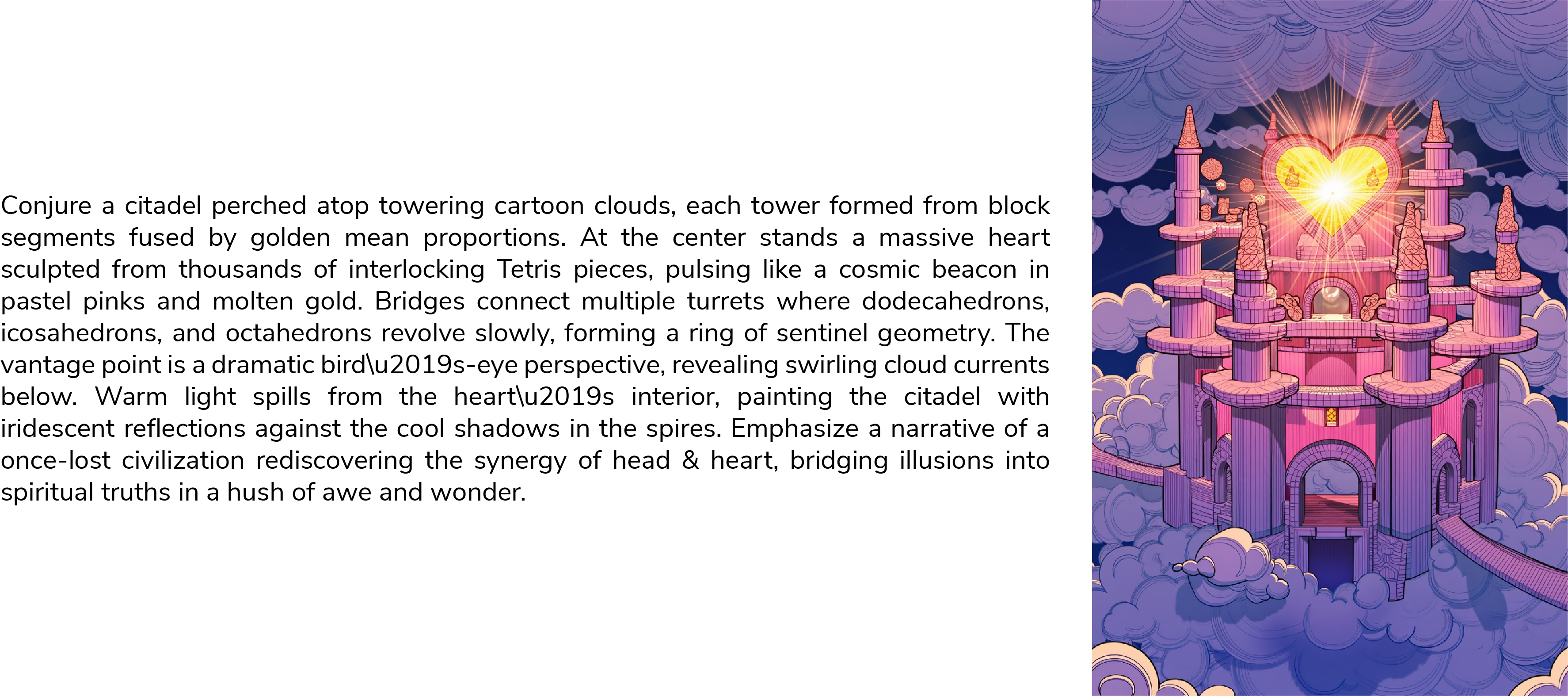}
    \\
    \includegraphics[width=\linewidth]{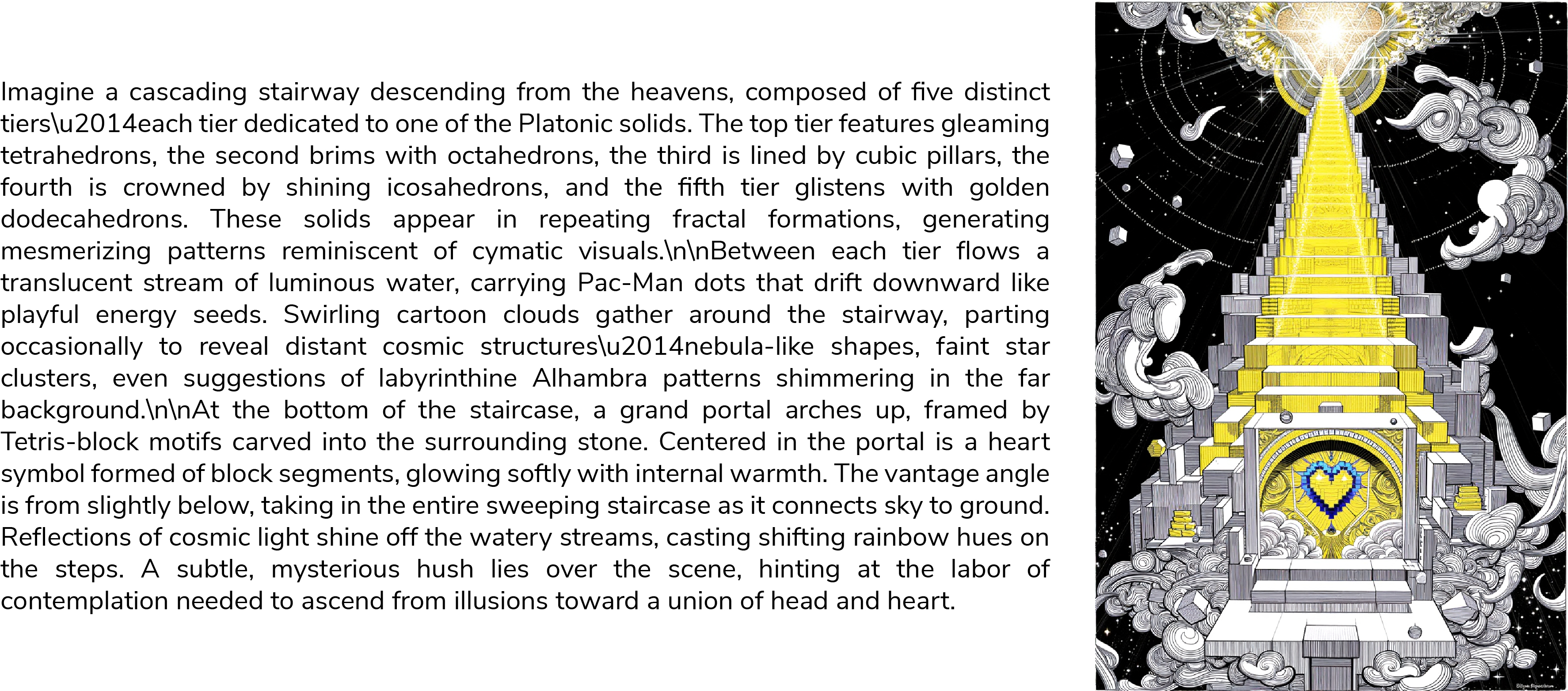}
    \\
    \caption{Examples of prompts and their corresponding outputs provided by BLK-Conceptor}.
    \label{fig:blk_conceptor_outputs}
\end{figure*}

\begin{figure*}
    \centering
    \includegraphics[width=\linewidth]{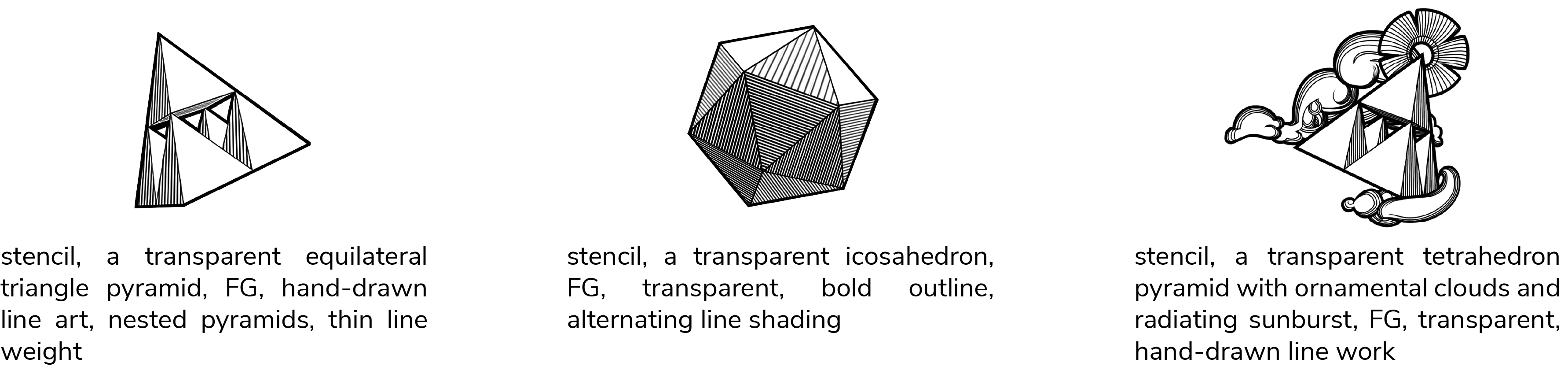}
    \\
    \includegraphics[width=\linewidth]{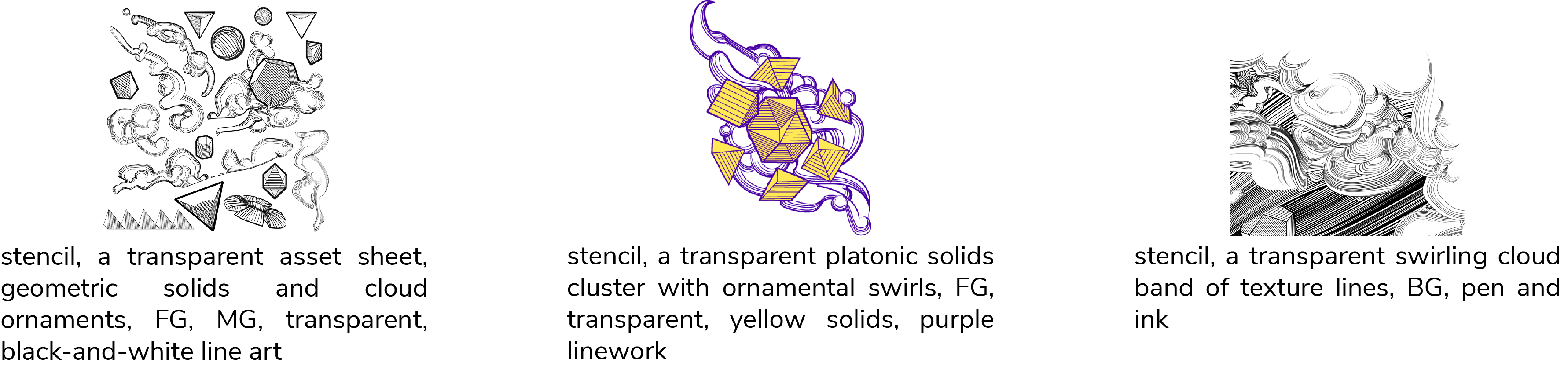}
    \\
    \includegraphics[width=\linewidth]{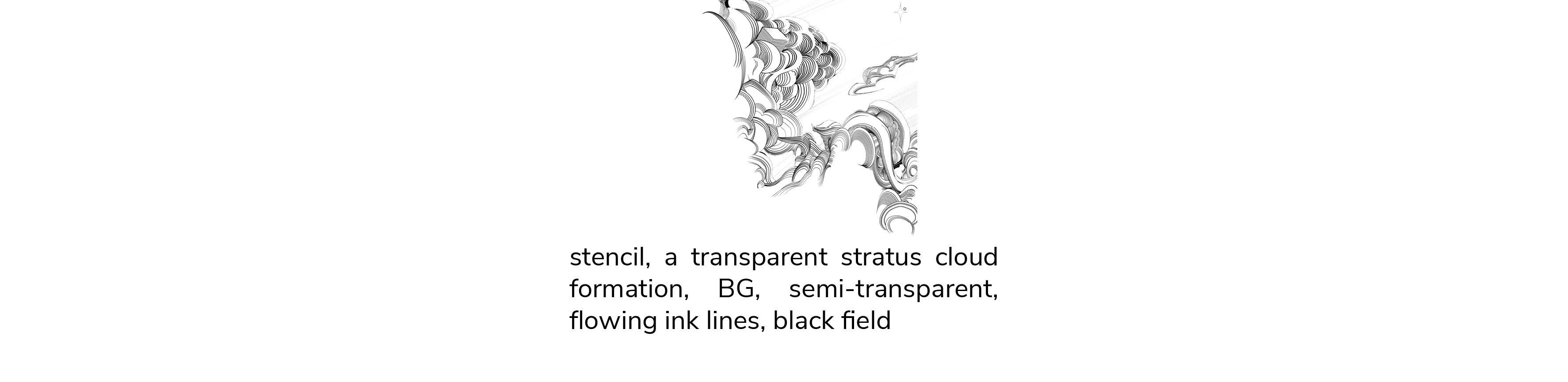}
    \\
    \caption{Input training data (PNGs and captions) for BLK-Stencil}.
    \label{fig:blk_stencil_inputs}
\end{figure*}

\begin{figure*}
    \centering
    \includegraphics[width=\linewidth]{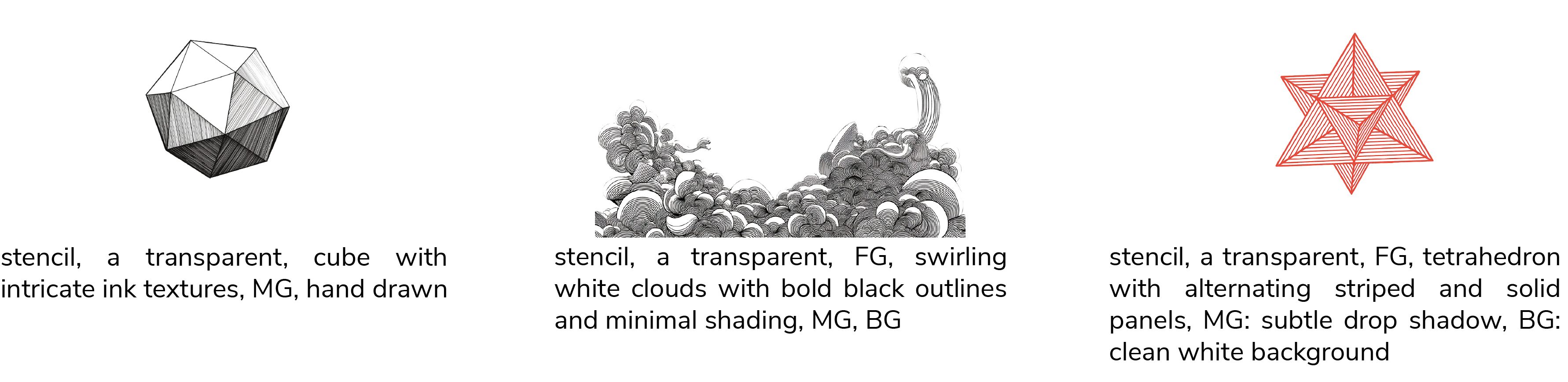}
    \\
    \includegraphics[width=\linewidth]{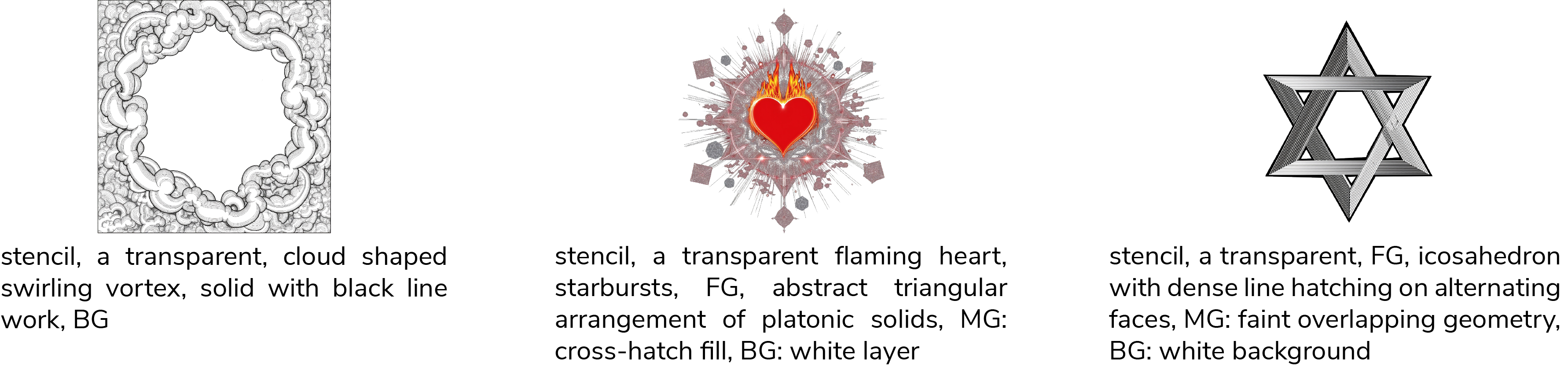}
    \\
    \includegraphics[width=\linewidth]{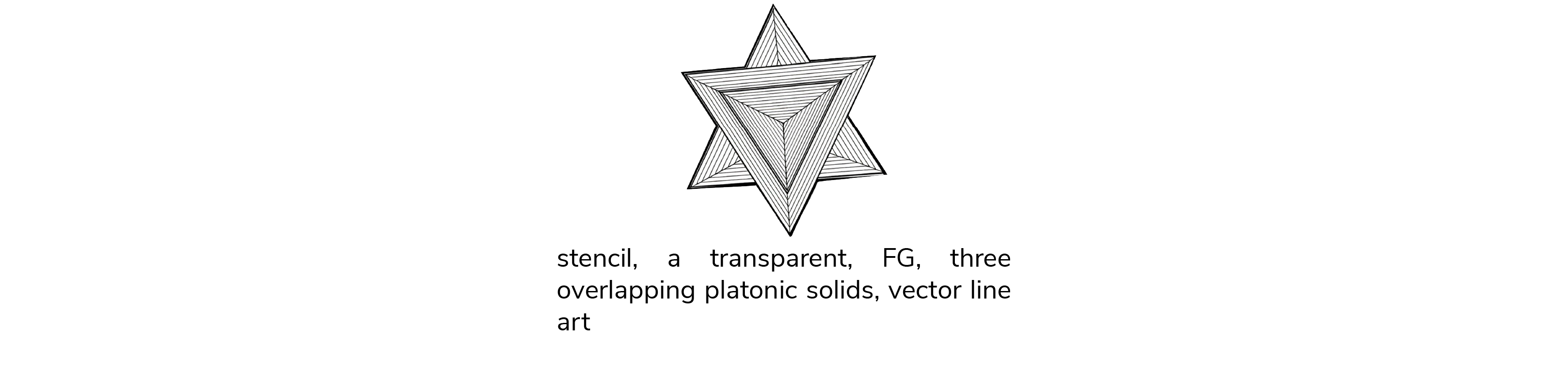}
    \\
    \caption{Examples of prompts and their corresponding outputs provided by BLK-Stencil}.
    \label{fig:blk_stencil_outputs}
\end{figure*}


\end{document}